\documentstyle[twocolumn,epsf]{IEEEtran}
\def\BibTeX{{\rm B\kern-.05em{\sc i\kern-.025em b}\kern-.08em
            T\kern-.1667em\lower.7ex\hbox{E}\kern-.125emX}}
\def\lesssim{\ \raise.3ex\hbox{$<$}\kern-0.8em\lower.7ex\hbox{$\sim$}\ }
\def\gesim{\ \raise.3ex\hbox{$>$}\kern-0.8em\lower.7ex\hbox{$\sim$}\ }

\newcommand{\ve}[1]{\mbox{\scriptsize \boldmath {$#1$}}}
\newcommand{\VE}[1]{\mbox{\boldmath {$#1$}}}

\input epsf

\begin{document}
\title{Theory of Photoemission-type Experiment in the BCS-BEC Crossover Regime of a Superfluid Fermi Gas}
\author{\vskip 1em
{\Large Ryota Watanabe, Shunji Tsuchiya, Yoji Ohashi}\thanks{Ryota Watanabe is with Department of Physics, Keio University, Yokohama, Japan (E-mail:rwatanab@rk.phys.keio.ac.jp). Shunji Tsuchiya is with Department of Physics, Tokyo University of Science, Tokyo, Japan, and JST(CREST), Saitama, Japan. Yoji Ohashi is with Department of Physics, Keio University, Yokohama, Japan, and JST(CREST), Saitama, Japan.
}
\vskip 1em
}
\maketitle
\begin{abstract}
We theoretically investigate the recent photoemission-type experiment on $^{40}$K Fermi gases done by JILA group. Including pairing fluctuations within a strong-coupling $T$-matrix theory, as well as effects of a harmonic trap within the local density approximation, we calculate photoemission spectra in the BCS (Bardeen-Cooper-Schrieffer)-BEC (Bose-Einstein condensation) crossover region. We show that the energy resolution of the current photoemission experiment is enough to detect the pseudogap phenomenon. We also show how the pseudogap in single-particle excitations continuously changes into the superfluid gap, as one decreases the temperature below the superfluid phase transition temperature. Our results would be useful for the study of single-particle properties of ultracold Fermi gases in the BCS-BEC crossover.
\end{abstract}
\section{Introduction} \label{section introduction}
\PARstart{T}{he} recent photoemission-type experiment done by JILA group\cite{STEWART,GAEBLER} is a very powerful method to examine single-particle properties of ultracold Fermi gases.
This Fermi system has the unique property that the strength of a pairing interaction can be tuned by adjusting the threshold energy of a Feshbach resonance\cite{GIORGINI,BLOCH}.
Using this, one can study from the weak-coupling Bardeen-Cooper-Schrieffer (BCS)-type superfluid to the Bose-Einstein condensation (BEC) of tightly bound molecules in a unified manner\cite{GIORGINI,BLOCH,EAGLES,LEGGETT,NSR,SADEMELO,Timmermans,Holland,OHASHI1}.
The continuous change from the weak-coupling BCS regime to the strong-coupling BEC regime is frequently referred to as the BCS-BEC crossover phenomenon, which is one of the most exciting topics in cold atom physics\cite{GIORGINI,BLOCH}. In particular, the so-called pseudogap phenomenon has recently attracted much attention\cite{Perali,PERALI,PIERI,Haussmann,Chen,TSUCHIYA1,TSUCHIYA2,WATANABE,HU}, because this strong-coupling phenomenon is also considered as a key to clarify the mechanism of high-$T_{\rm c}$ cuprates\cite{FISCHER}. Since the pseudogap appears in single-particle excitation spectra, the photoemission-type experiment developed by JILA group\cite{STEWART,GAEBLER} is very suitable for the study of this problem. Indeed, the observed excitation spectra in the crossover regime clearly deviate from those in the case of a free Fermi gas\cite{STEWART,GAEBLER}.
\par
In high-$T_{\rm c}$ cuprates, the mechanism of pseudogap has not been completely clarified yet, because of the complexity of the system (although the importance of pairing fluctuations has been pointed out\cite{FISCHER}). On the other hand, the origin of the pseudogap is well known in cold Fermi gases, namely, the formation of preformed Cooper pairs by strong pairing interaction. Thus, in the latter system, one can conveniently examine pseudogap physics originating from superfluid fluctuations, without any ambiguity. 
\par
In considering the photoemission-type experiment on cold Fermi gases, one should note the following two experimental situations. First, the system is always trapped in a harmonic potential\cite{Chen,HU}. As a result, pairing fluctuations are spatially inhomogeneous, which naturally leads to inhomogeneous pseudogap effects. For example, one may expect the situation that, while the pseudogap is remarkable in the trap center, such an effect is weak around the edge of the trap. Since the current photoemission experiment does not have spatial resolution\cite{STEWART}, it only gives spatially averaged data. Thus, it is an interesting problem how the pseudogap effect can be seen in such spatially averaged spectra. Second, the observed photoemission spectrum is always affected by finite energy resolution\cite{Chen}. Because of this, the pseudogap structure in the spectrum is expected to be smeared to some extent.
Thus, it is an important problem whether or not the current experiment with finite energy resolution ($D\sim 0.2\varepsilon_{\rm F}$, where $\varepsilon_{\rm F}$ is the Fermi energy) is enough to detect the interesting pseudogap phenomenon.
\par
Between the above two important issues, we have examined the first one in a previous paper\cite{TSUCHIYA1}. Including pairing fluctuations within a strong-coupling $T$-matrix theory, as well as effects of a harmonic trap within the local density approximation (LDA), we showed how the pseudogap appears in the spatially averaged photoemission spectra at the superfluid transition temperature $T_{\rm c}$. In this paper, we extend our previous work\cite{TSUCHIYA1} to include the experimental energy resolution. We show that the recent photoemission measurements on $^{40}$K Fermi gases\cite{STEWART,GAEBLER} really detect the pseudogap. We further extend this work to the superfluid state below $T_{\rm c}$. We clarify how the pseudogap continuously changes into the superfluid gap below $T_{\rm c}$. Far below $T_{\rm c}$, the single-particle spectral weight is shown to exhibit a double peak structure, originating from quantum fluctuations and superfluid gap.
\par
The outline of this paper is as follows. In Sec. II, we present our formulation. We explain how to include strong pairing fluctuations, effects of a harmonic trap, as well as finite energy resolution. In Sec. III, we show our numerical results on photoemission spectra. Throughout this paper, we take $\hbar=k_{\rm B}=1$. 

\section{Formalism}
We consider a two-component Femi gas, described by pseudo spin $\sigma=\uparrow,\downarrow$. In real cold Fermi gases, these pseudospins physically represent two atomic hyperfine states contributing to the formation of Cooper pairs. So far, all the current experiments on cold Fermi gases are using a broad Feshbach resonance to tune the strength of a pairing interaction\cite{STEWART,GAEBLER}. In this case, it is known that the pairing interaction may be simply treated by the ordinary BCS model\cite{GIORGINI}, as far as we consider the interesting BCS-BEC crossover physics. The model Hamiltonian is given by\cite{TSUCHIYA1,TSUCHIYA2,WATANABE,OHASHI}
\begin{equation}
\label{Hamiltonian}
H=\sum_{\ve p}\Psi^\dagger_{{\ve{p}}}[\xi_p \tau_3-\Delta\tau_1]\Psi_{{\ve p}}-U\sum_{\ve q}\rho_+({\VE q})\rho_-({\VE q}).
\end{equation}
Here, $\Psi_{\ve p}^\dagger=(c_{\ve p\uparrow}^\dagger,c_{-\ve p\downarrow})$ is the two-component Nambu field, where $c_{\ve p\sigma}$ is the annihilation operator of a Fermi atom with pseudospin $\sigma$. $\tau_j$ $(j=1,2,3)$ the Pauli matrices acting on the particle-hole space. $\xi_{\ve p}=\varepsilon_{\ve p}-\mu$ is the kinetic energy $\varepsilon_{\ve p}$ of a Fermi atom, measured from the chemical potential $\mu$. The pairing interaction is described by $-U$, which is assumed to be a tunable parameter. In cold atom physics, the strength of an interaction is usually measured in terms of the observable scattering length $a_s$. In the present BCS model, it is related to $-U$ as\cite{RANDERIA},
\begin{equation}
\frac{4\pi a_s}{m}=-\frac{U}{1-U\sum_{\ve p}^{\omega_c}\frac{1}{2\epsilon_p}},
\label{as}
\end{equation}
where $m$ is an atomic mass, and $\omega_c$ is a high-energy cutoff. In this scale, the weak-coupling BCS regime and the strong-coupling BEC regime are, respectively, given by $(k_{\rm F}a_s)^{-1}\lesssim -1$ and $(k_{\rm F}a_s)^{-1}\gesim1$ (where $k_{\rm F}$ is the Fermi momentum). The intermediate coupling region, $-1\lesssim (k_{\rm F}a_s)^{-1}\lesssim 1$, is the crossover region.
\par
The superfluid phase is characterized by the superfluid order parameter, $\Delta=\sum_{\ve p}\langle c_{-\ve p \downarrow}c_{\ve p \uparrow}\rangle$ (which is taken to be real, and is proportional to the $\tau_1$-component in Eq. (\ref{Hamiltonian}). In the last term of Eq.(\ref{Hamiltonian}), $\rho_\pm({\VE q})\equiv[\rho_1({\VE q})\pm i\rho_2({\VE q})]/2$ in Eq. (\ref{Hamiltonian}) involves the generalized density operators $\rho_j({\VE q})=\sum_{\ve p}\Psi^\dagger_{{\ve p}+{\ve q}/2}\tau_j\Psi_{{\ve p}-{\ve q}/2}$ ($j=1,2$), describing amplitude ($j=1$) and phase ($j=2$) fluctuations of the order parameter. 
\par
We treat the interaction term $-U$ in Eq. (\ref{Hamiltonian}) within the $T$-matrix approximation\cite{Perali,PERALI,PIERI,TSUCHIYA1,TSUCHIYA2,WATANABE}. For this purpose, we introduce the $2\times2$-matrix single-particle thermal Green's function, 
\begin{equation}
G_{\ve p}(i\omega_n)=\frac{1}{G_{\ve p}^0(i\omega_n)^{-1}-\Sigma_{\ve p}(i\omega_n,r)}.
\end{equation}
Here, $G_{\ve p}^0(i\omega_n)^{-1}\equiv i\omega_n-\xi_p\tau_3+\Delta\tau_1$ is the mean-field Green's function, where $\omega_n$ is the Fermi Matsubara frequency. The $2\times2$-matrix self-energy $\Sigma_{\ve p}(i\omega_n)$, involving effects of pairing fluctuations within the $T$-matrix approximation, has the form\cite{WATANABE}
\begin{eqnarray}
&&\Sigma_{\ve p}(i\omega_n)
\nonumber\\
&=&-T\sum_{\ve q,\nu_n}
\sum_{s,s'=\pm}\Gamma_{\ve q}^{s s'}(i\nu_n)
\tau_{-s}G_{\ve p+\ve q}^0(i\omega_n+i\nu_n)\tau_{-s'},
\nonumber\\
\label{self-energy}
\end{eqnarray}
where $\tau_\pm=\tau_1\pm i\tau_2$, and $\nu_n$ is the Bose Matsubara frequency. The scattering matrix in the Cooper channel $\Gamma_{\ve q}^{ss^\prime}(i\nu_n)$ is given by\cite{WATANABE}
\begin{eqnarray}
&&\left(
\begin{array}{cc}
\Gamma_{\ve q}^{+-}(i\nu_n)&
\Gamma_{\ve q}^{++}(i\nu_n)\\
\Gamma_{\ve q}^{--}(i\nu_n)&
\Gamma_{\ve q}^{-+}(i\nu_n)\\
\end{array}
\right)
\nonumber\\
&=&
-U
\left[
1+U
\left(
\begin{array}{cc}
\Pi_{\ve q}^{+-}(i\nu_n)&
\Pi_{\ve q}^{++}(i\nu_n)\\
\Pi_{\ve q}^{--}(i\nu_n)&
\Pi_{\ve q}^{-+}(i\nu_n)\\
\end{array}
\right)
\right]^{-1}.
\label{gamma}
\end{eqnarray}
Here,
\begin{eqnarray}
\label{eq.gamma}
&&\Pi_{\ve q}^{ss^\prime}(i\nu_n)
\nonumber
\\
&=&T\sum_{\ve p,\omega_n}{\mathrm Tr}
\Bigl[\tau_s G_{\ve p+\ve q/2}^0(i\omega_n+i\nu_n)\tau_{s'} G_{\ve p-\ve q/2}^0(i\omega_n)\Bigr]
\nonumber
\\ 
\end{eqnarray}
is the lowest order of the pair-correlation function in terms of the interaction $-U$.
\par
Experimentally, since a Fermi gas is always trapped in a harmonic potential $V(r)=m\omega_{\rm tr}^2r^2/2$\cite{note}, we include this inhomogeneous effect within the local density approximation (LDA), which is simply achieved by replacing the chemical potential $\mu$ by $\mu(r)=\mu-V_{\rm trap}(r)$. The single-particle Green's function and superfluid order parameter then have spatial dependences as $G_{\bf p}(i\omega_n,r)$ and $\Delta(r)$, respectively.
\par
The LDA photoemission spectrum $I({\VE p},\Omega)$ in a trapped Fermi gas is given by\cite{TSUCHIYA2}
\begin{equation}
\label{PES}
I({\VE p},\Omega)=\alpha \int_0^\infty r^2drp^2
A_{\ve p}(\xi_p(r)-\Omega,r)f(\xi_p(r)-\Omega),
\end{equation}
where $\xi_p(r)=\varepsilon_p-\mu(r)$, $f(\omega)$ is the Fermi distribution function, and $\alpha$ is a constant factor. (The detailed expression of $\alpha$ is not necessary in the following discussions.) The single-particle spectral weight $A_{\ve p}(\xi_p(r)-\Omega,r)$ at $r$ is obtained from the analytic continuation of the LDA Green's function as
\begin{equation}
\label{SW}
A_{\ve p}(\omega,r)=-\frac{1}{\pi}\textrm{Im}G_{\ve p}(i\omega_n\to\omega_+=\omega+i\delta,r)|_{11}.
\end{equation}
\par
As mentioned previously, the observed photoemission spectra in Ref. \cite{STEWART} are affected by experimental energy resolution. Incorporating this situation into Eq. (\ref{SW}), we have
\begin{equation}
{\bar I}({\VE p},\Omega)=\frac{1}{\sqrt{\pi}D}\int_{-\infty}^\infty dz I({\VE p},\Omega) e^{-\frac{(z-\Omega)^2}{D^2}}.
\label{SWE}
\end{equation}
For the value of energy resolution $D$, we employ the experimental value $D\simeq 0.2\varepsilon_{\rm F}$\cite{STEWART}. Equation (\ref{SWE}) is directly related to the occupied single-particle spectral weight $S({\VE p},\omega)$ as\cite{TSUCHIYA2}
\begin{equation}
S({\VE p},\omega)=
{1 \over \alpha}{\bar I}({\VE p},\Omega\to\xi_p-\omega).
\label{SWE2}
\end{equation}
In the simplest non-interacting Fermi gas at $T=0$, $S({\VE p},\omega)$ has a $\delta$-functional peak line along the free-particle dispersion $\omega=\xi_p$ when $\omega<0$. The spectral weight in the positive energy region vanishes due to the vanishing Fermi distribution function $f(\omega>0)=0$ at $T=0$.
\par
To calculate Eq. (\ref{SWE}), we first determine the superfluid order parameter $\Delta(r)$ and chemical potential $\mu$, which is achieved by solving the gap equation
\begin{equation}
\label{BCSEQ}
1={4\pi a_s \over m}\sum_{\ve p}\left(\frac{\textrm{tanh}\frac{E_p(r)}{2T}}{2E_p(r)}-\frac{1}{2\epsilon_p}\right)=0,
\end{equation}
(where $E_{\ve p}(r)=\sqrt{\xi_p(r)^2+\Delta(r)^2}$ is the Bogoliubov single-particle excitation spectrum), together with the LDA number equation for Fermi atoms (within the $T$-matrix approximation), 
\begin{equation}
\label{NUMBEREQ}
N=\int_0^\infty 4\pi r^2dr 2T\sum_{\ve p,\omega_n}G_{\ve p}(i\omega_n,r)|_{11}e^{i\delta\omega_n}.
\end{equation}
The superfluid phase transition temperature $T_{\rm c}$ is obtained under the condition that the gap equation (\ref{BCSEQ}) is satisfied when $\Delta(r=0)=0$. Using the self-consistent solutions, we calculate the spectral weight in Eq. (\ref{SWE2}). 
\par

\begin{figure}[t]
\centering
\epsfxsize=.3\textwidth
\epsfbox{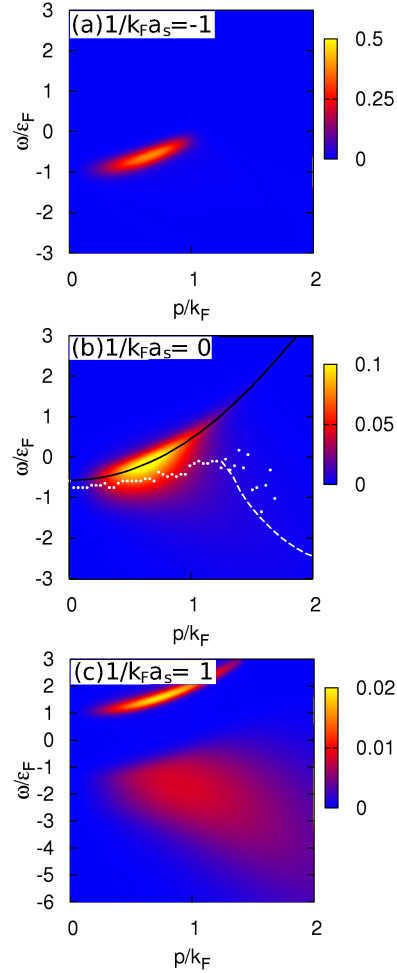}\\
\caption{Calculated intensity of occupied spectral weight $S({\bf p},\omega)$ in a trapped Fermi gas at $T_{\rm c}$. The bright color region shows high intensity. (a) $(k_{\rm F}a_s)^{-1}=-1$. (b) $(k_{\rm F}a_s)^{-1}=0$. (c) $(k_{\rm F}a_s)^{-1}=+1$. In panel (b), the black solid line shows the upper peak position of the spectrum, and white dashed line shows the lower peak positions of the spectrum. (Although the peak intensity of the lower line cannot be seen in panel (b), this is simply because its magnitude is much smaller than the dominant peak intensity around $p/k_{\rm F}\sim 0.5$.) The white dots represent the experimental data for the lower peak positions measured in Ref.\cite{STEWART}.}
\label{FIG1}
\end{figure}


\begin{figure}[t]
\centering
\epsfxsize=.3\textwidth
\epsfbox{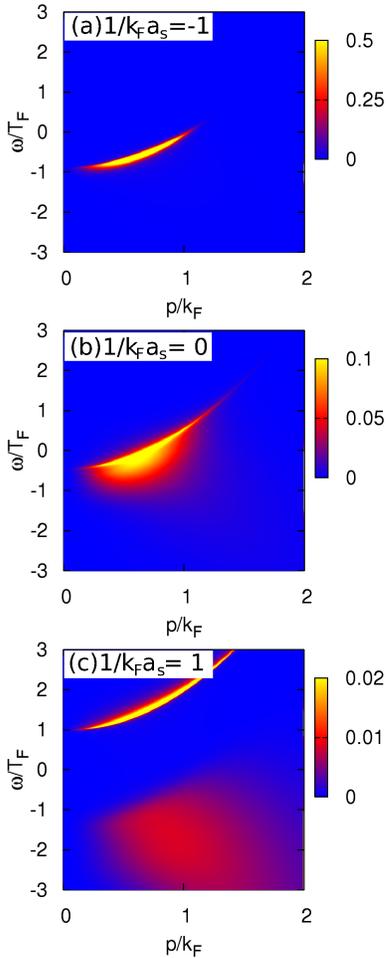}\\
\caption{Same plots as in Fig.\ref{FIG1}, in the case when the experimental energy resolution $D$ is ignored.}
\label{FIG2}
\end{figure}

\begin{figure}[t]
\centering
\epsfxsize=.3\textwidth
\epsfbox{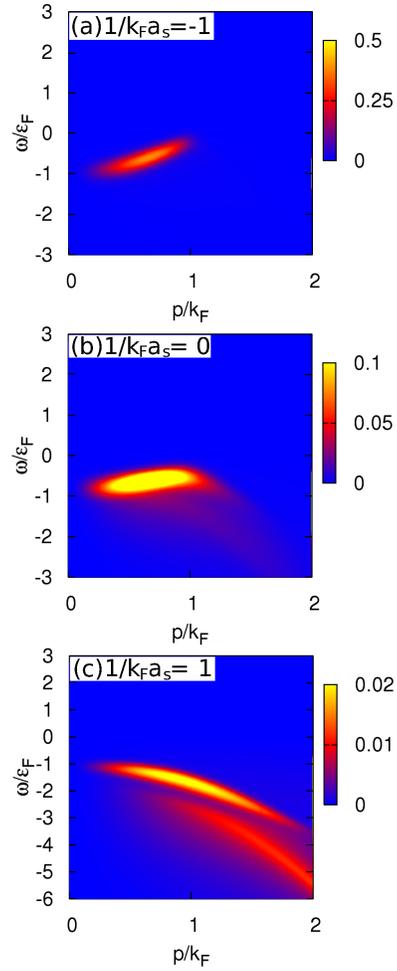}\\
\caption{Same plots as in Fig.\ref{FIG1}, for $T=0.1T_{\rm c}$ (superfluid phase).}
\label{FIG3}
\end{figure}


\section{Photoemission spectra with finite energy resolution in the BCS-BEC crossover}
\par
Figure \ref{FIG1} shows the occupied single-particle spectral weight $S({\VE p},\omega)$ at $T_{\rm c}$. Comparing this figure with the case with no energy resolution ($D=0$) shown in Fig.\ref{FIG2}, we find that the finite energy resolution ($D = 0.2\varepsilon_{\rm F}$) broadens the spectra. However, even in this case, one can still see the pseudogap effect in Fig.\ref{FIG1}. That is, starting from the weak-coupling BCS regime shown in Fig.\ref{FIG1}(a), we see that the spectral peak gradually deviates from the free particle dispersion, as the pairing interaction becomes strong. (See Fig.\ref{FIG1}(b).) In the strong-coupling BEC regime (Fig.\ref{FIG1}(c)), the single-particle excitation spectrum splits into an upper sharp particle branch ($\omega>0$) and lower broad hole branch ($\omega<0$). Since the superfluid order parameter is absent at $T_{\rm c}$, the pseudogap structure in Fig.\ref{FIG1}(c) purely comes from strong pairing fluctuations. As shown in Fig.\ref{FIG1}(b), the calculated lower peak line (white dashed line in the figure) agrees well with the recent experiment on a $^{40}$K Fermi gas done by JILA group. As discussed in our previous papers\cite{TSUCHIYA1,TSUCHIYA2,WATANABE}, this lower peak is characteristic of the pseudogap effect, originating from a particle-hole coupling induced by pairing fluctuations. (Note that, when the pairing interaction is absent, the photoemission spectrum only has the upper peak (black solid line in Fig. \ref{FIG1}(b)) line along the dispersion of a free atoms. These results naturally lead to the conclusion that the energy resolution $D\simeq 0.2\varepsilon_{\rm F}$ at the current stage of photoemission experiment is enough to detect the pseudogap phenomena in the BCS-BEC crossover regime of cold Fermi gases.
\par
We now proceed to the superfluid phase below $T_c$. Far below $T_{\rm c}$, since thermal pairing fluctuations are almost absent, the occupied spectral weight $S({\VE p}, \omega)$ in the positive energy region almost vanishes, as shown in Fig.\ref{FIG3}. In the ordinary mean-field BCS theory for a uniform Fermi superfluid, the single-particle Bogoliubov excitation spectrum in the hole branch is given by
\begin{equation}
\omega=-E_p=-\sqrt{\xi_p^2+\Delta^2}.
\label{hole}
\end{equation}
Although thermal fluctuations are almost absent far below $T_{\rm c}$, Fig.\ref{FIG3} shows that the peak positions of the occupied spectral weight still deviate from the expected mean-field result in Eq. (\ref{hole}). In particular, one sees two branches in panels (b) and (c).
\par
\begin{figure}[!htp]
\centering
\epsfxsize=.3\textwidth
\epsfbox{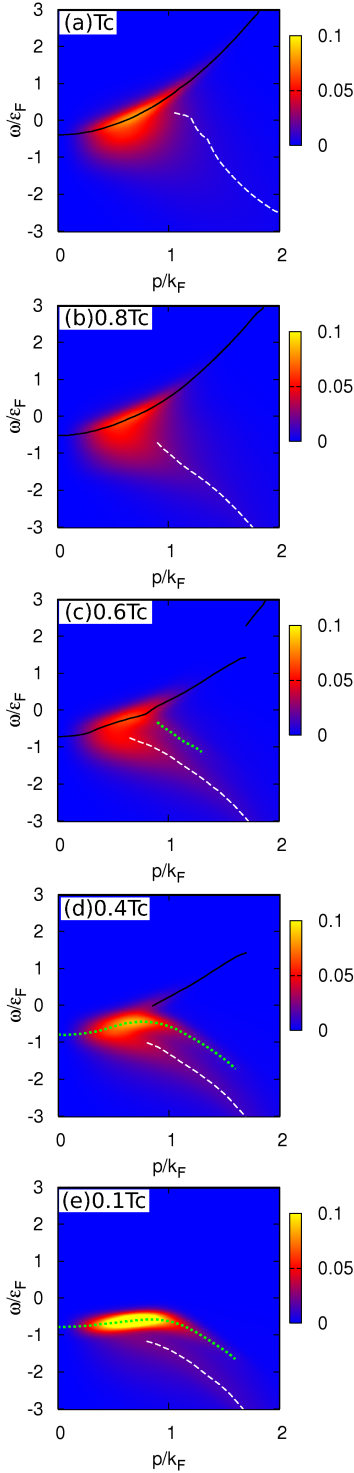}\\
\caption{Calculated temperature dependence of occupied single-particle spectral weight $S({\bf p},\omega)$. We set $1/(k_{\rm F}a_s)=0.15$. The lines shown in this figure show peak positions of the spectral weight.
The calculated chemical potential $\mu$ are (a)$0.33\epsilon_{\rm F}$, (b)$0.43\epsilon_{\rm F}$, (c)$0.47\epsilon_{\rm F}$, (d)$0.56\epsilon_{\rm F}$, and (e)$0.60\epsilon_{\rm F}$.
}
\label{FIG4}
\end{figure}

To see the origin of the appearance of two branches in Figs.\ref{FIG3}(b) and (c), we show in Fig.\ref{FIG4} the temperature dependence of the occupied spectral weight in the crossover region ($1/k_{\rm F}a_s=0.15$). As one decreases the temperature below $T_{\rm c}$, panel (b) indicates that the lower peak line (dashed line) is pushed down by the development of superfluid order parameter. At lower temperatures shown in panel (c), the other peak line (green dotted line) appears, which gradually reduces to the dispersion of hole Bogoliubov excitations in Eq. (\ref{hole}), as shown in panels (d) and (e). That is, between the two branches in Figs.\ref{FIG3}(b) and (c), the lower one is related to the pseudogap at $T_{\rm c}$, and the upper one originates from the ordinary BCS excitation gap. As mentioned previously, since thermal fluctuations are almost absent far below $T_{\rm c}$, the lower peak at $T\ll T_{\rm c}$ is considered to be also related to quantum fluctuations.  
\par

\begin{figure}[h]
\centering
\epsfxsize=.3\textwidth
\epsfbox{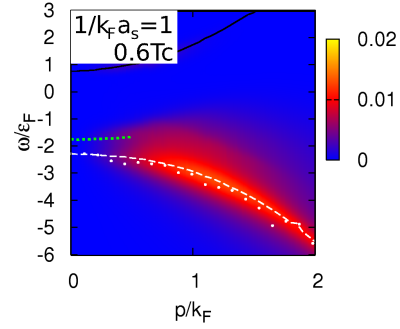}\\
\caption{Calculated occupied single-particle spectral weight $S({\bf p},\omega)$ in the BEC regime at $(k_{\rm F}a_s)^{-1}=1$. We take at $T=0.6T_{\rm c}$. For clarity, we also draw three lines at the peak positions of the spectral weight. The white dots are experimental data at $(k_{\rm F}a_s)^{-1}=1$\cite{STEWART}.
}
\label{FIG5}
\end{figure}

In a previous paper\cite{TSUCHIYA2}, we showed that, in the BEC regime, the lower peak energies in $S({\VE p},\omega)$ calculated at $T_{\rm c}$ are slightly larger than the experimental data by JILA group\cite{STEWART}. In this regard, we note that it is difficult to accurately determine the temperature in cold Fermi gases. Thus, since the lower peak line is pushed down in the superfluid phase (See Fig.\ref{FIG4}.), a possible idea to resolve this discrepancy is that the experiment in the BEC regime was actually done in the superfluid phase below $T_{\rm c}$. Indeed, when we take $T=0.6T_{\rm c}<T_{\rm c}$, our theoretical result in the BEC regime ($1/(k_{\rm F}a_s)=+1$) well explains the observed peak energies, as shown in Fig.\ref{FIG5}. Although we need further studies about this problem, our result shows that the existence of a finite superfluid order parameter is a possible idea to explain the photoemission spectrum observed in the BEC regime.

\begin{figure}[h]
\centering
\epsfxsize=.4\textwidth
\epsfbox{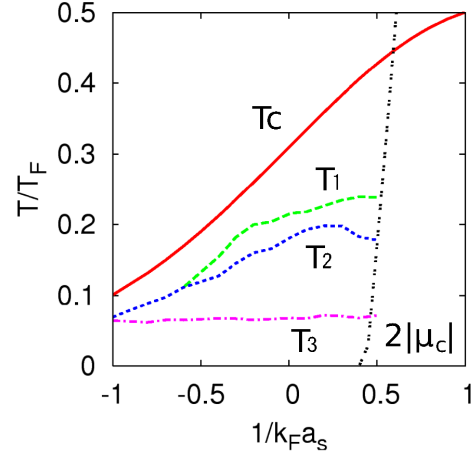}\\
\caption{Characteristic temperatures obtained from the occupied single-particle spectral weight $S({\bf p},\omega)$. Below $T_1$, the peak line associated with the BCS excitation gap appears in $S({\bf p},\omega)$. The left edge of this peak line reaches $p=0$ at $T_2$. Below $T_2$, this line looks similar to the dispersion of hole Bogoliubov excitations in Eq. (\ref{hole}). Below $T_3$, the spectral weight no longer has a finite weight in the positive energy region within our numerical accuracy. In this figure, we also plot $T=2|\mu(T_{\rm c})|$ when $\mu<0$. Since $2|\mu(T_{\rm c})|$ gives the binding energy of a two-body bound state, it physically means the characteristic temperature, above which two-bound states dissociate into atoms thermally. Thus, the right side of this line may be viewed as the region of a gas of tightly bound molecules, rather than strongly correlated Fermi atoms.
}
\label{FIG6}
\end{figure}

\par
To summarize the behavior of $S({\VE p},\omega)$ in the BCS-BEC crossover below $T_{\rm c}$, we introduce three characteristic temperatures shown in Fig.\ref{FIG6}. Just below $T_{\rm c}$, although the system is in the superfluid phase, one cannot see any superfluid effect in $S({\VE p},\omega)$. Below a certain temperature $T_1$, a peak structure corresponding to the BCS excitation gap appears in the spectrum. One may also introduce another characteristic temperature $T_2$, below which this peak line is well described by the Bogoliubov excitation spectrum in Eq. (\ref{hole}). Since thermal excitations are suppressed far below $T_{\rm c}$, $S({\VE p},\omega)$ becomes absent in the positive energy region $\omega>0$. The characteristic temperature $T_3$ is defined as the temperature when this situation is realized. 
\par
In Fig.\ref{FIG6}, we also plot the binding energy of a two-body bound molecule, give by $E_g=2|\mu(T_{\rm c})|$\cite{TSUCHIYA1,TSUCHIYA2,WATANABE}. When the temperature is lower than $E_g$, thermal dissociations of molecules are suppressed, so that the system is close to a molecular Bose gas formed by a two-body effect, rather than a strongly correlated Fermi gas. 
\par

\section{Summary}\label{section:conclusions}

To summarize, we have theoretically investigated the recently realized photoemission-type experiment on ultracold Fermi gases. Our theory takes into account pairing fluctuations within the $T$-matrix, as well as effects of a harmonic trap within the local density approximation. To include the experimental situation, we have also taken into account the energy resolution $D=0.2\varepsilon_{\rm F}$ in calculating the photoemission spectrum.
\par
At $T_{\rm c}$, we showed that, although the energy resolution broadens the photoemission spectra, we can still see the pseudogap effect in the BCS-BEC crossover region. Below $T_{\rm c}$, we also showed how the pseudogapped spectral weight continuously changes into the spectral weight with a finite BCS superfluid gap. To summarize the temperature dependence of spectral weight below $T_{\rm c}$, we have introduced three characteristic temperatures, $T_1$, $T_2$ and $T_3$. Since the photoemission-type experiment is one of the most powerful methods to observe single-particle properties of cold Fermi gases, our results would be useful for the understanding of strong-coupling effects in the BCS-BEC crossover regime of this system.

\subsection*{Acknowledgments}
We would like to thank J. P. Gaebler and D. S. Jin for providing us with their experimental data. We also thank S. Watabe, D. Inotani, and T. Kashimura for fruitful discussions. This work was supported by Global COE Program ``High-Level Global Cooperation for Leading-Edge Platform on Access Spaces (C12)'', as well as the Japan Society for the Promotion of Science.



\begin{thebibliography}{}
\bibitem{STEWART}J. T. Stewart, C. A. Regal, and D. S. Jin, Nature (London) \textbf{454}, 744 (2008).
\bibitem{GAEBLER}J. P. Gaebler, J. T. Stewart, T. E. Drake, D. S. Jin, A. Perali, P. Pieri, and G. C. Strinati, Nature Phys. \textbf{6}, 569 (2010)
\bibitem{GIORGINI}S. Giorgini, S. Pitaevskii, and S. Stringari, Rev. Mod. Phys. \textbf{80}, 1215 (2008).
\bibitem{BLOCH}I. Bloch, J. Dalibard, and W. Zwerger, Rev. Mod. Phys. \textbf{80}, 885 (2008).
\bibitem{EAGLES}D. M. Eagles, Phys. Rev. \textbf{186}, 456 (1969).
\bibitem{LEGGETT}A. J. Leggett, \textit{Modern Trends in the Theory of Condensed Matter} (Springer, Berlin, 1960).
\bibitem{NSR}P. Nozi\`eres and S. Schmitt-Rink, J. Low. Temp. Phys. \textbf{59}, 195 (1985).
\bibitem{SADEMELO}C. A. R. Sa de Melo, M. Randeria, and J. R. Engelbrecht, Phys. Rev. Lett, \textbf{71}, 3202 (1993).
\bibitem{Timmermans}E. Timmermans, K. Furuya, P. W. Milonni, and A. K. Kerman, Phys. Lett. A \textbf{285}, 228 (2001).
\bibitem{Holland}M. Holland, S. J. J. M. F. Kokkelmans, M. L. Chiofalo, and R. Walser, Phys. Rev. Lett \textbf{87}, 120406 (2001).
\bibitem{OHASHI1}Y. Ohashi, and A. Griffin, Phys. Rev. A \textbf{67}, 063612 (2003).
\bibitem{Perali}A. Perali, P. Pieri, G. C. Strinati, and C. Castellani, Phys. Rev. B \textbf{66}, 024510 (2002).
\bibitem{PERALI}A. Perali, P. Pieri, L. Pisani, and G. C. Strinati, Phys. Rev. Lett. \textbf{92}, 220404 (2004).
\bibitem{PIERI}P. Pieri, L. Pisani, and G. C. Strinati, Phys. Rev. B \textbf{70}, 094508 (2004).
\bibitem{Haussmann}R. Haussmann, M. Punk, and W. Zwerger, Phys. Rev. A \textbf{80}, 0636122 (2009).
\bibitem{Chen}Q. J. Chen and K. Levin, Phys. Rev. Lett. \textbf{102}, 190402 (2009).
\bibitem{TSUCHIYA1}S. Tsuchiya, R. Watanabe, and Y. Ohashi, Phys. Rev. A \textbf{80}, 033613 (2009).
\bibitem{TSUCHIYA2}S. Tsuchiya, R. Watanabe, and Y. Ohashi, Phys. Rev. A \textbf{82}, 033629 (2010).
\bibitem{WATANABE}R. Watanabe, S. Tsuchiya, and Y. Ohashi, Phys. Rev. A \textbf{82}, 043630 (2010).
\bibitem{HU}H. Hu, X. J. Liu, P. D. Drummond, and H. Dong, Phys. Rev. Lett. \textbf{104}, 240407 (2010).
\bibitem{FISCHER}O. Fischer, M. Kugler, I. Maggio-Aprile, C. Berthod, and C. Renner, Rev. Mod. Phys. \textbf{79}, 353 (2007)
\bibitem{RANDERIA}M. Randeria, in \textit{Bose-Einstein Condensation}, edited by A. Griffin, D. W. Snoke, and S. Stringari (Cambridge University Press, New York 1995), p. 355.
\bibitem{OHASHI}Y. Ohashi, and S. Takada, J. Phys. Soc. Jpn. \textbf{66}, 2437 (1997).
\bibitem{note} The anisotropy of a trap potential $V_{\rm trap}=\sum_{j=x,y,z}m\omega_j^2r_j^2/2$ is irrelevant within LDA in the sense that the trap potential can be always mapped onto the isotropic one by an appropriate scale transformation. Thus, in this paper, we only consider the case of isotropic trap.
\end{thebibliography}
\end{document}